\documentclass[aps,prl, showkeys]{revtex4}

\usepackage{amssymb}
\usepackage{amsmath}
\usepackage{graphicx}
\usepackage{float}
\usepackage{multirow}

\begin{document}

%************************************************************************************************
\title{Correlation and collective behaviour in Adler-type locally coupled oscillators at the edge of chaos}
%************************************************************************************************

% Use letters for affiliations, numbers to show equal authorship (if applicable) and to indicate the corresponding author
\author{Ernesto \surname{Estevez-Rams}}
\email{estevez@fisica.uh.cu}
\affiliation{Facultad de F\'isica-Instituto de Ciencias y Tecnolog\'ias de Materiales, Universidad de la Habana, San L\'azaro y L, CP 10400 Habana, Cuba}

\author{K. \surname{Garcia-Medina}}
\affiliation{Facultad de F\'isica, Universidad de la Habana, San L\'azaro y L, CP 10400 Habana, Cuba}

\author{B. \surname{Arag\'on-Fern\'andez}}
\affiliation{Universidad de las Ciencias Inform\'aticas. Carretera a San Antonio, Boyeros. La Habana. Cuba }


\begin{abstract}
Dynamical systems can be analyzed as computational devices capable of performing information processing. In coupled oscillators, enlarged capabilities are expected when the set of units is formed by subsets with collective behaviour within them but weak correlation between the subsets. A system of non-linear oscillators weakly coupled in the phase approximation is studied. The informational distance maps show different regimes of collective behaviour, ranging from independent, local, and global, as the control parameters of the system are changed. This rich set of behaviours happens despite the simple nature of the model used. Complex hierarchies between oscillators can be identified near the edge of chaos region in the informational distance dendrograms. We identify the emergence of local collective regimes, further corroborated by the spatiotemporal maps. Correlation diagrams also exhibit non-trivial dependence for the system at the edge of chaos. 
\end{abstract}

%\PACS{05.45.Tp,02.50.Ga,02.90.+p}

\date{\today}
\maketitle

\textbf{Introduction}. Non-linear coupled oscillators have been used to model several natural phenomena \cite{guckenheimer83,mosekilde02}. They exhibit a wide range of behaviours from trivial periodic dynamics to chaotic, including instances with all the fingerprints of a complex system \cite{alonso17,alonso18,estevez18,burylko22,song23}.

Defining complexity remains an elusive task despite the considerable literature on the subject. One of many possible approaches is to consider the emergence of global dynamics in systems composed of many interacting elements, each simple when isolated \cite{estevez18}. However, everyone seems to agree that lasting correlations along the components of the system are a necessary feature \cite{grassberger86}. Such correlations can be recognized by the emergence of patterns persistent in time, yet it still allows for unpredictability or randomness. The interaction between randomness and structured dynamics makes optimal prediction of the system time evolution a formidable task \cite{zambella88} but always better than mere guessing.

Related, but not quite a measure of complexity, is Kolmogorov or algorithmic complexity \cite{kolmogorov65} $K(s)$. The idea is to measure randomness by the length of the smallest algorithm $s^*$  that, in a Universal Turing Machine (UTM), can reproduce the system's output $s$ ($K(s)=|s^*|$). Appealing to a UTM makes this measure absolute up to a constant or, at most, a slowing increasing function of the system size \cite{vitanyi93}. Finding the smallest algorithm is undecidable, a consequence of the halting problem \cite{vitanyi93}, but it is unnecessary in many cases. For most evaluations, it suffices to learn how the length of the shortest algorithm scales with system size, a rate equivalent to the Kolmogorov-Sinai entropy or entropy density $h$ of a system \cite{zurek89}. $h$ measures the unpredictability of the system's output or the ability to generate new information \cite{cover06}. 

Kolmogorov complexity is, therefore, a measure of disorder, not complexity. However, it is the starting point for the formal derivation of other magnitudes that measure correlations and are better suited for quantifying complexity in some way. One of these magnitudes is the effective complexity ($E$) introduced by Grassberger \cite{grassberger86}, also known as excess entropy \cite{crutchfield03}. $E$ measures the amount of correlation in a system at all scales. Formally, it is the mutual information between the two halves of a bi-infinite one-dimensional output \cite{crutchfield03}.

From a computational point of view, several related but different approaches can be taken to recognize complexity. The degree of complexity of a dynamical system is related to the class of computational tasks that it can perform \cite{wolfram02}. In this approach, the system is seen as a computational device capable of performing effective computations. Computation, as used in this context, must not be considered an anthropocentric concept. Effective computation in nature can include but is not limited to human understandable algorithmic perspective, such as that involved in computer programming as understood in computer science. However, a broader scope is to consider that any system capable of generating, modifying, transmitting, and storing information in some way is a computational device.

The increasing evolution of neural networks has accommodated the use of computational devices in the sense explained above, where they are not viewed as algorithmic programable devices but as entities with information processing and storage ability \cite{appeltant10}. This greatly increases the interest in finding and characterizing such systems.

The degree of universality in the ability to compute goes through answering the question of how adaptable the system is to a change in the initial input conditions. Suppose a system is indifferent to the initial conditions, not changing its behaviour as the input dramatically changes. In that case, it has a limited computational capacity, i.e., it always performs the same task regardless of the initial stimulus. On the other extreme, if it has a widely varying behaviour with small changes in the initial conditions, its reliability as a computational device is questionable, as its robustness against noise is rather weak. Between those two extreme cases, the most universal type of computing occurs. 

Another important consideration is the ability of the system to correlate its components in such a way as to enhance its collective strength beyond some trivial manner. It is well known that it is in the collective dynamics of neurons that thought emerges \cite{gerstner14}. Also, in terms of memory, the system must be able to retain, even in a modified way, some of the initial or the generated information during its evolution. Again, memory is known to be a result of the collective behaviour of neurons.

It is clear by now that measuring non-trivial correlations between the units of a system is an important step toward understanding it as a computational device.

The fact that non-linear coupled oscillators have been used to model to some degree the brain activity \cite{gerstner14,ashwin16} is a statement of the kind of complex behaviour such systems can simulate. Nevertheless, it is unclear a priori which specific instance of coupled oscillators is capable of what degree of computation. 

A rich set of behaviours has been found in a simple model of locally coupled non-linear oscillators \cite{alonso17,estevez18,estevez23}. By changing the control parameters of the system, the set of coupled oscillators can exhibit trivial, chaotic, and complex behaviour. More interesting is the claim that enhanced computation capability can be seen right in the border between the chaotic and the complex region, the so-called edge of chaos (EOC) region \cite{estevez19}. 

The edge of chaos hypothesis refers to the emergence, in non-linear systems of different nature \cite{chua05,estevez18,estevez19,garcia23b}, of a surge in self-organization along the transition between order and chaos, a region well-defined and fundamentally unstable, with a stark competition between order and disorder \cite{langton90}. Some authors even link this hypothesis to the so-called ''criticality'' under which the brain seems to operate \cite{alexander96}, or evolution and life itself \cite{bilder14,lewin99}. 

In this contribution, we use informational distance \cite{li04} to study the correlation between oscillators. The informational distance  is based on how innovative two sequences $s$ and $p$ are, one with respect to the other in a symmetrical way, in terms of pattern production, when the other is taken into account 
\begin{equation}
d=\frac{\max\{K(s|p), K(p|s)\}}{\max\{K(s), K(p)\}},\label{eq:d}
\end{equation}
where $K(x|y)$ is the Kolmogorov complexity of $x$ conditioned in knowing $y$ (the length of the shortest algorithm reproducing $x$, if the shortes algorithm reproducing $y$, is given). The normalized distance $d$ is between $0$ and $1$.

This distance is not a Hamming-type distance because it does not measure the number of different bits between the sequences but how far they are in algorithmic space.

As defined in (\ref{eq:d}), the distance is uncomputable for the same reason as the algorithmic complexity. Therefore, an estimation procedure is needed. There have been several approaches to such estimation, but we will be following \cite{estevez15} (See the methods section for details). 

As already stated, entropy density 
\[
 h=\lim_{| s | \rightarrow \infty}\frac{H[s]}{| s |},
\]
where $H[s]$ is the Shannon block entropy of the string $s$, and effective complexity measure 
 \[
E = \sum_{L=1}^{\infty}(h(L)-h),
\]
with $h(L)=H[s(1, L)]-H[s(1, L-1)]$, will be used as a measure of randomness and the levels of correlation at different scales in the system, respectively.

$h(L)$ can be taken as a finite-size estimate of $h$, and therefore, $h(L)-h$ quantifies how much we are overestimating the actual randomness of the sequence when only correlations only at scales smaller than $L$ are accounted for. There are several interpretations for the effective measure complexity \cite{grassberger86,crutchfield03}, and the reader is encouraged to read the relevant literature. It will suffice to say that it measures the amount of patterns in a signal when noise has been accounted for. In this sense, knowing that is the mutual information between past and present, equivalently, between two halves of a bi-infinite sequence, it measures the persistence of structural features in a sequence at all time (length) scale. As with any other single magnitude, it can not capture the full scope of what we understand by complexity, but together with the other entropic magnitudes; it moves in the direction of characterizing it quantitatively.

First, we show the emergence of local collective behaviour in the EOC region. The local nature of the collective behaviour indicates a complex scenario where different groups of oscillators interact. Second, we show that the local collective behaviour emerging at the EOC changes as the system parameters change towards the interior of the needle region, gaining in scope until nearly global informational coherence is achieved.

It is also interesting to notice that, despite being such a simple model, dynamics expected in more involved models already emerge in this simple setup.

We should expect very limited computational capacity when each oscillator is independent of the others, and no collective behaviour is seen. Conversely, the computational capacity should also be reduced if all oscillators are synchronized in a phase-lock regime. For rich processing capability, one should expect that some local correlation between the oscillator must be present while global synchronization is not attained. The interaction between those local subsets and collective behaviour can lead to higher information processing capacity.

%***************************************************************************************************************************************************
\textbf{The model.} Let us consider a system of locally coupled Adler-type oscillators. Each oscillator in isolation is governed by a law of the type \cite{adler73}
\begin{equation}
\frac{d\theta}{dt}=\dot{\theta}=\omega+\gamma \cos \theta,\label{eq:adler}
\end{equation}
where $\omega$ is the intrinsic frequency of the oscillator, and $\gamma$ determines the strength of the self-feedback. Equation (\ref{eq:adler}) has been the subject of research in the past as an oscillator with an injection of a sinusoidal signal \cite{li10}, as given by the $\theta$-dependent feedback term in the right-hand side, that separates it from the harmonic oscillators. The feedback is a control mechanism to phase-lock the oscillator synchronized to a frequency source. Phase locking refers to the output phase of the oscillator being related to the input phase, in this case, the injection signal. Adler systems, as described by equation (\ref{eq:adler}), is just but one of a family of oscillators with injection feedback that can be realized electronically \cite{li10}.

Analytical treatment of equation (\ref{eq:adler}) results in critical points $\dot{\theta}=0$, given by $\cos(\theta^*)=-\omega/\gamma$, which can only happen if $|\omega/\gamma|\le 1$. Two points of different stability exist if the strict inequality happens, and one marginally stable critical point when equality happens ($|\omega/\gamma|=1$). For $\omega=\gamma$, oscillations starting with a lower $\theta$ phase will be attracted to the critical point, while those starting above the critical value $\theta^*$ will be repelled. For $|\omega/\gamma|<1$, a saddle-node bifurcation leads to two critical points, one stable and one unstable. In the asymptotic $t\rightarrow \infty$ time, the oscillator settles into $\sin \theta=\pm \sqrt{1-(\omega/\gamma)^2}$.  When $|\omega/\gamma|>1$, the phase speed $\dot{\theta}$ has a $\theta$ interval where a slowdown occurs, whose magnitude is related to the value of $|\omega/\gamma|-1$. The solution is periodic for all time values with frequency $\omega_a=1/2\sqrt{\omega^2-\gamma^2}$.

For the coupled oscillator system, the following local coupling introduced by Alonso \cite{alonso17} will be used
\begin{equation}
\frac{d\theta_i}{dt}=\dot{\theta_i}=\omega+\gamma \cos \theta_i+(-1)^i\left(\cos\theta_{i-1}+\cos \theta_{i+1}\right),\label{eq:alonso}
\end{equation}
which has been thoroughly studied \cite{alonso17,estevez18,estevez23} . The coupling is taken to depend on the absolute value of neighbouring phases, and its alternating sign answers to the local balancing of positive and negative injections. This choice of coupling differs from the usual Kuramoto coupling based on phase difference and not absolute values. There is no priory justification for such coupling other than its possible realization as an electronic circuitry. The fact that it has shown complex structured behaviour in some regions of the parameter space resembling other systems with computational abilities points to its potential usefulness \cite{estevez23}. Considering an even total number of oscillators guarantees global balancing of interactions. The overall effect of this balance can be thought of as the absence of a net injection to the system, thus avoiding the presence of a drift velocity for the entire system in the phase circle.  

\begin{figure}[!t]
\centering
\includegraphics*[scale=0.65]{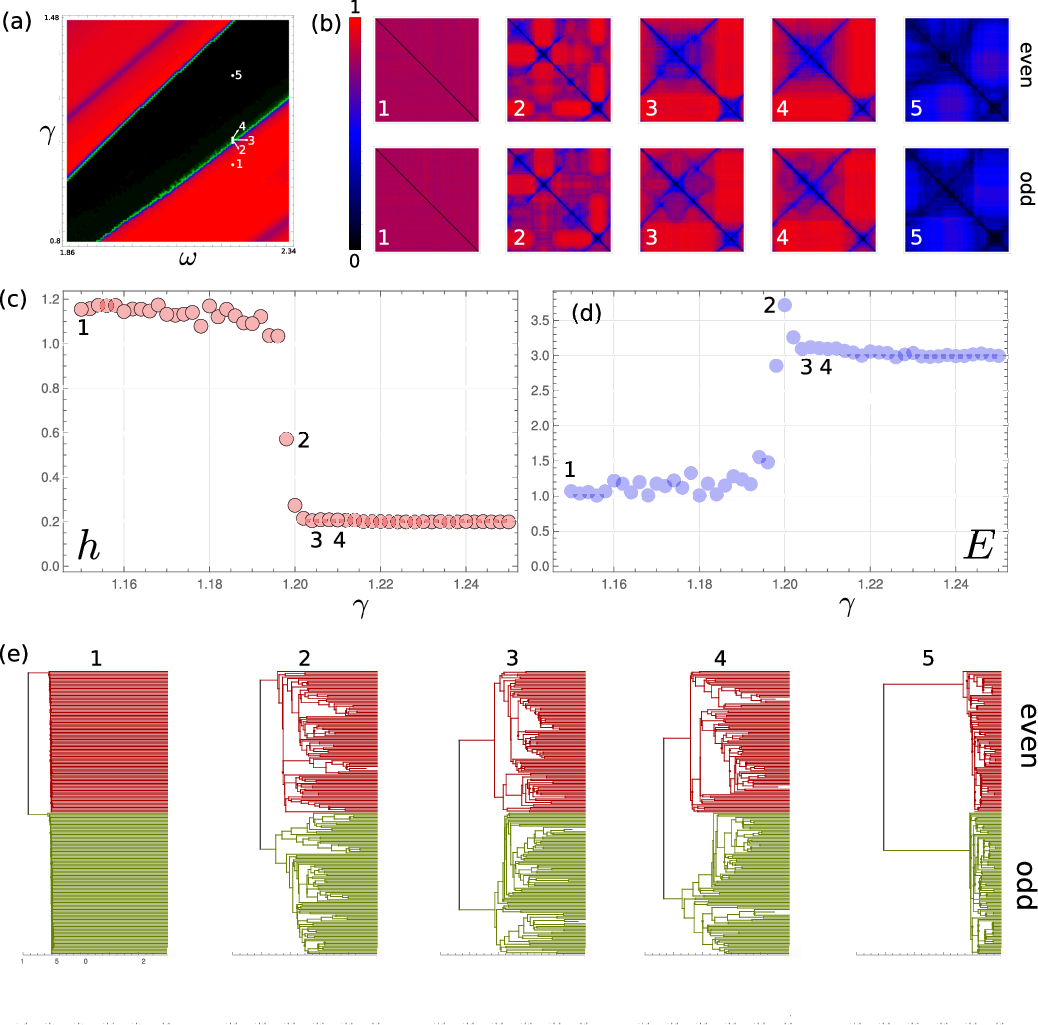}
\caption{(a) Entropy density map in parameter space for the locally coupled oscillators of equation (\ref{eq:alonso}). The needle region is the low $h$ (black) region. Five points have been chosen, all with $\omega=2.225$, and (1) $\gamma=1.150$, (2) $\gamma=1.200$, (3) $\gamma=1.205$, (4) $\gamma=1.210$ and (5) $\gamma=1.350$. The high $h$ (red) region is in the chaotic regime, where point (1) is taken. Point (2) is at the edge of chaos boundary. Points (3) and (4) are in the needle region close to the EOC, while point (5) is well within the needle region. (b) The distance matrix for a 200 oscillator configuration in all five points described. The distance map is shown separately for even and odd oscillators. (c) The entropy density $h$ as a function of $\gamma$ for $\omega=2.225$, as $\gamma$ increase the systems leaves the chaotic region and gets into the needle region, the boundary is seen as a sudden drop in the $h$ value. (d) The effective complexity measure $E$ as a function of $\gamma$, the same conditions as in (c) applies. A surge in the $E$ value can be seen at the EOC boundary. (e) The dendrogram built from the corresponding distance matrix. In the chaotic region (point 1), oscillators are independent between them, and the dendrogram is a trivial tree with a shallow deepness; within the needle region (point 5), the information distance between oscillators has dropped significantly within oscillators of one type (same sign), a more global collective behaviour has emerged. In the intermediate points (2, 3 and 4), at or near the EOC, a complex dendrogram shows the emergence of local collective behaviours with non-trivial hierarchies. At the same time, the global coupling is not achieved.
}\label{fig:1}
\end{figure}

Equation (\ref{eq:alonso}) splits the oscillators into two classes, given by the sign of the neighbourhood feedback. The parameter space of this system shows regions of periodic, chaotic, and complex behaviour. It is especially noteworthy that it has been reported, in one of the borders between the chaotic and the needle complex region (figure \ref{fig:1}a), a surge in magnitudes such as $E$, for some initial conditions \cite{estevez19} (See figure \ref{fig:1}c). In this edge of chaos (EOC) border, there is a drop in entropy density $h$ (figure \ref{fig:1}b) as the system enters the complex region, an expected behaviour as the system moves out of the chaotic regime. The reader can find further details in \cite{alonso17,estevez18,estevez23}.

%**************************************************************************************************************************************************
\textbf{Results and discussion.} In the region where a drop of the $h$ values and a surge in $E$ values is observed (Figure \ref{fig:1}c-d), the correlation among oscillators should exhibit a more involved picture. This is what is shown in figure \ref{fig:1}b, where the informational distance matrix or map between oscillators is shown at the EOC (point 2: $\omega=2.225$, $\gamma=1.200$) compared to neighbouring points in the parameter space ($\omega=2.225$, $\gamma=1.150, 1.205, 1.210, 1.350$, points 1, 3, 4 and 5). 

\begin{figure}[!t]
\centering
\includegraphics*[scale=0.85]{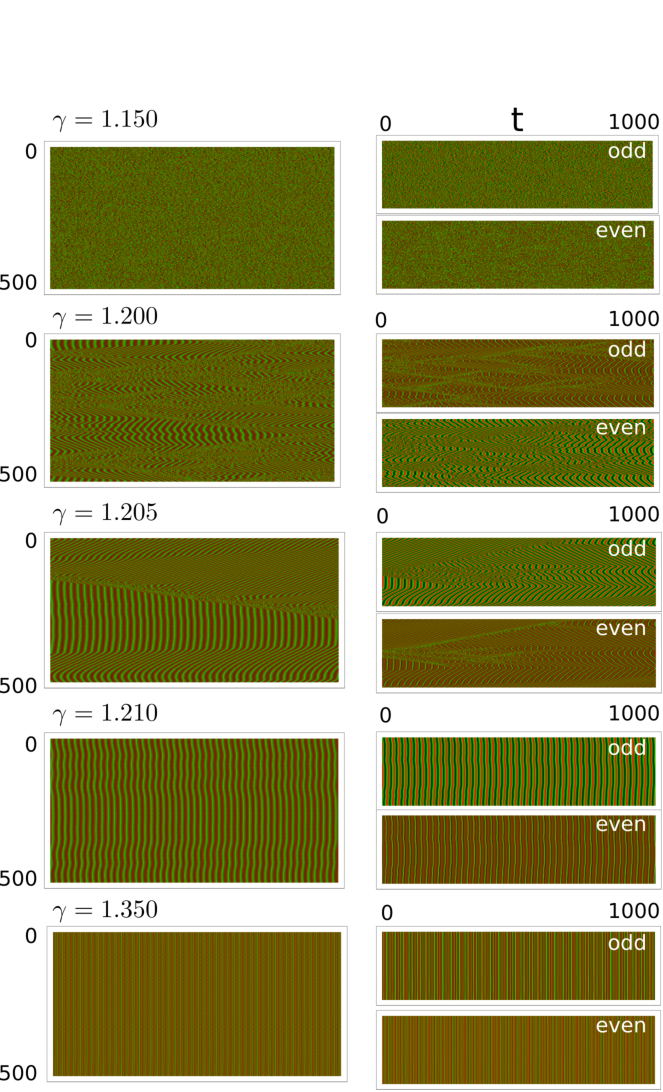}
\caption{Spatio-temporal maps for the five points depicted in figure \ref{fig:1}. The activity $\sin \theta$ values are shown as colours ranging from black for $0$, green for $1/2$, and red for $1$. Time runs horizontally while the oscillator index runs vertically. On the left is the whole spatiotemporal map; on the right, the spatiotemporal map is decomposed into odd and even oscillators. For $\gamma=1.1500$, a salt and pepper contrast shows that each oscillator behaves independently of the others in the chaotic region. No pattern can be discerned. Collective behaviour as time-persistent patterns emerges from EOC at $\gamma=1.200$ and becomes more dominant as $\gamma$ increases to $1.35$ where total global coherence can be seen. At EOC, there is a balance between pattern and incoherent regions in the oscillator dynamics.
}\label{fig:2}
\end{figure}

Figure \ref{fig:1}b-above shows the distance map among the even oscillators, while figure \ref{fig:1}b-below corresponds to the odd oscillators. The split between odd and even oscillators was made to take into account that each class has a different sign feedback, given by the $(-1)^i$ term,  as shown in equation (\ref{eq:alonso}). In the chaotic regime (point 1, $\gamma=1.150$), the oscillators are essentially free in their behaviour, and no collective dynamics are found. This is shown in the distance map and the spatiotemporal plot of figure \ref{fig:2}, where the evolution of the oscillators' phase at discrete time steps is shown as a colour map. This is in clear contrast with the distance matrix of the EOC region (point 2, $\gamma=1.200$), where non-trivial correlations between oscillators emerge and can be seen in the patterns formed in the distance map. The spatiotemporal diagram of figure \ref{fig:2} also shows the formation of patterns persistent in time and travelling along the oscillators: collective behaviour has emerged. Yet, the presence of incoherent boundaries between the coherent oscillators points to the local character of this coherency. Another interesting feature is that collective regions do not necessarily come from the oscillators' initial (random) phase values, as coherent regions can emerge or die during time evolution: the emergence and persistence of patterns is a finite time event.

For points 3 and 4, near the EOC but more into the needle region, the distance map shows less patterning (figure \ref{fig:1}b), while well into the needle region (point 5), decoherence seems to be lost for any oscillator and coherence has become global. The spatiotemporal maps support this picture; as the system moves from the EOC region into the needle region, coherence becomes increasingly global and persistence in time becomes longer, which can be seen in figure \ref{fig:2} for $\gamma=1.205, 1.210$, and $\gamma=1.350$ it already shows a complete pattern driven spatiotemporal behaviour.

The dendrogram from the distance matrix (Figure \ref{fig:1}e) further emphasizes the above analysis. Whereas for the chaotic region (point 1), the corresponding tree is shallow, with all pairs of oscillators showing a maximum distance between them, well within the needle region (point 5), the dendrogram shows that within a given oscillator class (odd or even, according to the sign of $(-1)^i$ in equation (\ref{eq:alonso}), the distance between oscillators has decreased significantly. This decrease in distance points to global correlations between the oscillators. In contrast, at or near the EOC (points $2$, $3$, $4$), the dendrogram shows complex hierarchies among the oscillators, with several levels of deepness in the tree within a class of oscillators. Local communities of oscillators can be identified as sub-trees in the dendrogram for these three points. Enhancement is not a property just of the EOC at point $2$, although it seems to be an optimal value as shown by the surge in $E$. Suboptimal improvement, with respect to the neighbouring regions, is still a feature for points $3$ and $4$, as can be seen by the fact that the dendrogram (figure \ref{fig:1}e) at points $2$, $3$ and $4$, shows no qualitative difference and just slight quantitative ones, (e.g., the mean distance between odd oscillators shown in figure \ref{fig:1} are, respectively, $0.797$, $0.826$ and $0.843$. In contrast, for point $5$ is $0.391$). Fine-tuning the control parameters to a very tight interval of control parameter values is unnecessary. Of course, this could be related to the finite number of oscillators far from any thermodynamic limit, yet infinity is never achieved in real systems.    

Simulations were performed with several initial phase values, random and non-random. The distance matrix shows an evident dependence on the initial condition, a feature of an adaptable computational device (See supplementary material). 

Figure \ref{fig:3} shows the plots of the average value of $\Delta h_{ij}=|h_i-h_{i+2j}|$ and $\Delta E_{ij}=|E_i-E_{i+2j}|$, averaged over $i$ as a function of $j$, for odd and even oscillators. For independent oscillators, both magnitudes should be constant with $j$, the observed behaviour for the chaotic region $\gamma=1.150$. A constant behaviour should be observed for globally coupled oscillators, such as all oscillators in phase. This is almost the functional dependence seen for $\gamma=1.210, 1.350$. The fact that there is a weak dependence with $j$  between the oscillators indicates that complete phase lock has not been attained. For locally collective behaviour, with decoupling between these regions of collective oscillators, a more involved functional dependence of $\Delta h_{ij}$ and $\Delta E_{ij}$ should be expected, and this is the case for $\gamma=1.205$, even more so for the EOC point, $\gamma=1.200$. For the EOC plot, it must be noted that the non-trivial function of both magnitudes includes monotonicity changes and plateaus, indicating some correlation between oscillators that are not necessarily close to each other. $\Delta h_{ij}$ and $\Delta E_{ij}$ vs $j$ is, for $\gamma=1.200$, a result of the balance between pattern or collective behaviour and unpredictability or lack of correlation.

 At EOC, the value of $E$ for the oscillators is between $4$ and $10$, while the entropy density $h$ is between $0.1$ and $0.5$. Both intervals are intermediate compared to the points outside this region. $h$ interval at EOC is lower than in the chaotic region and larger than inside the needle region, while the $E$ interval at EOC is above the interval for the chaotic regime and below the needle region (See supplementary material).

 %***************************************************************************************************************************************
\textbf{Conclusions.} The picture that emerges from the analysis is that, indeed, at the EOC, the surge of $E$ with the drop of $h$ is witnessing the emergence of complex dynamics. The observed balance between structuring and unpredictability is precisely the fingerprint of complexity. The non-trivial local collective behaviour discovered by the patterns in the distance map, the dendrogram, and the spatiotemporal map indicates this complex dynamic. The reported result argues for enhanced computational power in the EOC region. However, this optimal improvement can still be suboptimal for neighbouring points on the side of the needle region. 

Current research is being done to see if similar findings hold for other systems of non-linear coupled oscillators. In other coupled oscillator systems, different coupling forms and topologies are being considered \cite{garcia23a,garcia23b}, and results will be reported elsewhere. It should be emphasized that the procedure presented is of general application and is not limited to the type of systems reported. As long as a time series output can be achieved from a given model, entropic measures can be estimated, and distance metrics can follow.

 %*****************************************************************************************************************************************
\textbf{Methods.}  In all cases, integration of the system defined by equation (\ref{eq:alonso}) was performed using a fourth-order Runge-Kutta method as implemented on GNU Scientific Library (GSL). In random instances, the results were compared to the solution given by the numerical solver of Mathematica \cite{Mathematica}; in all cases, both solutions were identical within the numerical error. Also, different integration steps were used until further refinement of the time step did not change the reported result.  

Since all entropic markers introduced are defined for discrete-state systems, a discretization of the original continuous activity is needed. For the sake of simplicity, a binary alphabet was chosen. Furthermore, given the bounded nature of the $sin(\theta)$ function, the average activity at any given time becomes a convenient choice for the threshold value. Binarization was thus carried, assigning a $1$ state to all oscillators with activity higher than the mean value and $0$ otherwise. This has been done before in similar systems \cite{estevez18,estevez23}. The question of generating partition is one with no universal answer; one expects that the chosen discretization keeps sufficient information on the relevant features in the analyzed system. If the discretization proves to be successful in doing so, it can only be evaluated a posteriori. In any case, the thresholding procedure does not introduce artefacts in the data processing but merely reduces the amount of information carried by the data. 

Entropy density $h$ was estimated through a Lempel-Ziv factorization as explained in \cite{estevez19}. Briefly, Lempel-Ziv factorization \cite{lz76} creates an exhaustive history from the input by defining a factor every time an unseen substring happens while scanning the input string from left to right. 

The LZ76 complexity $C(s)$ of the sequence $s$ is defined as the number of factors in its exhaustive history.

Let the entropy rate be given by 
\begin{equation}\label{shannonh}
 h(s)= \lim_{N\rightarrow\infty}\frac{H[s(1,N)]}{N},
\end{equation}
where $H[s(1,N)]$ is the Shannon block entropy \cite{cover06} of the string $s(1,N)\in s$. Then, defining
\begin{equation}\label{lzh}
 c(s)=\frac{C(s)}{N/\log{N}}.
\end{equation}
Ziv \cite{ziv78} proved that, if $s$ is the output from an ergodic source, then
\begin{equation}\label{zivtheorem}
 \limsup_{N\rightarrow\infty}c(s)=h(s).
\end{equation}
This is the base of using $c(s)$ as an estimate of $h$ for $N\gg 1$. 

Equation (\ref{zivtheorem}) is valid in the infinite limit; in practical cases, equation (\ref{lzh}) is used as an estimate for the entropy density. Assuming an i.i.d. Gaussian distribution of the word length, the error estimate for the entropy density estimated by Lempel Ziv can be calculated from \cite{amigo06}
\begin{equation}
 \sigma=c^{3/2} \frac{\varsigma}{\sqrt{N \log N}},\label{eq:sigma}
\end{equation}
$\varsigma$ is the standard deviation of the word length. For a $10^4$ length sequence, which is the one used in this study, equation (\ref{eq:sigma}) gives an order of magnitude for the error bound around $10^{-2}$. Lesne et al. \cite{lesne09} have further discussed the estimation of error in the use of entropy estimators in short sequences. Using numerical simulations, they showed that Lempel-Ziv estimators of entropy density show good agreement even for sequences as short as $2\times 10^3$. As stated, sequence length will be at least $10^4$, and errors in estimating entropy density should be insignificant. 

From the entropy density, a magnitude related to the effective complexity can readily estimated by a random shuffle algorithm \cite{estevez19}. 

As Kolmogorov randomness is uncomputable, a practical alternative to equation (\ref{eq:d}) is needed \cite{estevez15}. $K(s|p)$ is roughly equal to $K(sp)-K(p)$ (there is a logarithmic correction term which is not significant for long enough sequences), and $sp$ denotes the concatenation of both strings. The informational distance can then be estimated from 
\begin{equation}
 d(s,p)=\frac{K(sp)-min \{ K(s), K(p) \} }{max\{ K(s), K(p)\}}.\label{eq:dnidk}
\end{equation}
If $s$ and $p$ have the same length, then we can cast equation (\ref{eq:dnidk}) in terms of the entropy density
\begin{equation}
 d(s,p)=\frac{h(sp)-min \{ h(s), h(p) \} }{max\{ h(s),h(p)\}}.\label{eq:dnidh}
\end{equation}
 The normalized LZ76 complexity is now used to estimate the entropy density,
\begin{equation}
 d(s,p)=\frac{c_{LZ}(sp)-min\{c_{LZ}(s), c_{LZ}(p)\}}{max\{c_{LZ}(s), c_{LZ}(p)\}}.\label{eq:dlz}
\end{equation}

For the distance matrix, simulations were performed for $200$ oscillators with circular boundary conditions, random initial phase values were taken, and $6000$ time steps, the first $1000$ time steps were dropped before the similarity distance was calculated. Dendrograms were built using the average intercluster dissimilarity as criteria from the distance matrix. The Dendrogram function of Wolfram Mathematica was used \cite{Mathematica}.

$5000$ oscillators were also used with periodic boundary conditions for the spatiotemporal maps. Again, the first $1000$ time steps were dropped, and the next $1000$ time steps are shown. The $\langle \Delta h_j \rangle$ and $\langle \Delta E_j \rangle$ plots were calculated for $500$ oscillators, as before the first $1000$ time steps were dropped.  

 %*************************************************************************************************************************
\textbf{Acknowledgment.} EER and KGM would like to thank Alexander von Humboldt Stiftung for financial support. EER would like to thank the MPI-PKS for a visiting grant. EER thanks H. Kantz at MPI-PKS for the in-depth discussion and the excellent working environment.

%\bibliographystyle{unsrt}
%\bibliography{adler_dm.bib}

\pagebreak

%\begin{table}[!htbp]
%\centering
%\caption{Distance statistics ($\omega=2.225$). $d_{min}$, $d_{max}$, $\langle d \rangle$, $
%\sigma$ stands for the minimum, maximum, mean, and standard deviation distance between oscillators. The distances reported are split between odd (inhibitory) and even (excitatory) oscillators.}
%\begin{tabular}{c@{\hspace{2em}}c@{\hspace{2em}}c@{\hspace{2em}}c@{\hspace{2em}}c@{\hspace{2em}}c}
%\toprule\\
%$\gamma$ & & $d_{min}$ & $d_{max}$ & $\langle d \rangle$ & $\sigma$ \\
%\hline\\
%\multirow{3}*{1.150}& odd & 0.787 & 0.873 & 0.831 & 0.01\\\\
% & even & 0.793 & 0.860 & 0.829 & 0.01 \\\hline\\
%\multirow{3}*{1.200}& odd & 0.119 & 0.957 & 0.797 & 0.1\\\\
% & even & 0.077 & 0.981 & 0.823 & 0.1 \\\hline\\
% \multirow{3}*{1.205}& odd & 0.048 & 0.971 & 0.826 & 0.1\\\\
% & even & 0.139 & 0.980 & 0.814 & 0.1 \\\hline\\
%  \multirow{3}*{1.210}& odd & 0.034 & 0.971 & 0.843 & 0.1\\\\
% & even & 0.080 & 0.979 & 0.829 & 0.1 \\\hline\\
%   \multirow{3}*{1.350}& odd & 0.040 & 0.600 & 0.391 & 0.1\\\\
% & even & 0.038 &  0.667& 0.401 & 0.1 \\
%\toprule
%\end{tabular}
%\end{table}

\begin{figure}[!t]
\centering
\includegraphics*[scale=0.8]{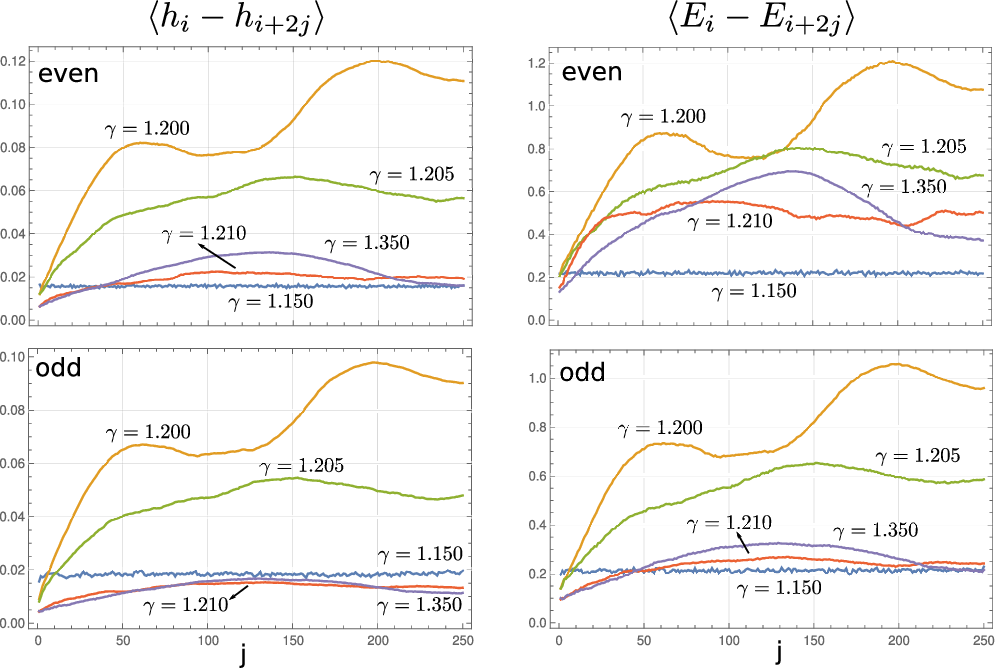}
\caption{Plot for the five described points, of the mean value of $\Delta h_{ij}=|h_i-h_{i+2j}|$ (left) and $\Delta E_{ij}=|E_i-E_{i+2j}|$ (right) as a function of $j$ (averaged over $i$). For independent oscillators, both magnitudes should not depend on $j$, the behaviour observed for $\gamma=1.150$. Global coherence with near-phase locking should also show a slow dependence with $j$ as seen for $\gamma=1.210, 1.350$. Local coherent regions and incoherent oscillators should show a non-trivial function of $j$ as seen for $\gamma=1.200, 1.205$.
}\label{fig:3}
\end{figure}

\end{document}